\documentclass{article}

\usepackage[preprint,%
nonatbib]{nips_2018}

\newcommand*{\published}{22 May 2018}

\newcommand*{\pdftitle}{Maximum-entropy and representative samples\\ of neuronal
  activity: a dilemma}

 \newcommand*{\pdfauthor}{P.G.L.  Porta Mana, V. Rostami, E. Torre, Y. Roudi}
\usepackage{newtxmath}
\usepackage[T1]{fontenc} 
\input{glyphtounicode} \pdfgentounicode=1
\usepackage[utf8]{inputenx}

\usepackage{textcomp}
\usepackage{amsmath}
\usepackage{mathtools}
\usepackage{amsxtra}

\usepackage[british]{babel}

\usepackage[autostyle=false,autopunct=false,english=british]{csquotes}
\setquotestyle{british}

\usepackage{amsthm}

\theoremstyle{remark}

\newtheoremstyle{innote}{\parsep}{\parsep}{\footnotesize}{}{}{}{0pt}{}
\theoremstyle{innote}

\usepackage{bm}

\DeclareFontFamily{U}{egreek}{\skewchar\font'177}%
\DeclareFontShape{U}{egreek}{m}{n}{<-6>s*[0.95]eurm5 <6-8>s*[0.95]eurm7 <8->s*[0.95]eurm10}{}%
\DeclareFontShape{U}{egreek}{m}{it}{<->s*[0.95]eurmo10}{}%
\DeclareFontShape{U}{egreek}{b}{n}{<-6>s*[0.95]eurb5 <6-8>s*[0.95]eurb7 <8->s*[0.95]eurb10}{}%
\DeclareFontShape{U}{egreek}{b}{it}{<->s*[0.95]eurbo10}{}%
\DeclareSymbolFont{egreeki}{U}{egreek}{m}{it}%
\SetSymbolFont{egreeki}{bold}{U}{egreek}{b}{it}
\DeclareSymbolFont{egreekr}{U}{egreek}{m}{n}%
\SetSymbolFont{egreekr}{bold}{U}{egreek}{b}{n}
\DeclareFontFamily{U}{egreekx}{\skewchar\font'177}
\DeclareFontShape{U}{egreekx}{m}{n}{%
       <-7.5>s*[0.9]euex7%
    <7.5-8.5>s*[0.9]euex8%
    <8.5-9.5>s*[0.9]euex9%
    <9.5->s*[0.9]euex10%
}{}
\DeclareSymbolFont{egreekx}{U}{egreekx}{m}{n}
\DeclareMathSymbol{\sumop}{\mathop}{egreekx}{"50}
\DeclareMathSymbol{\prodop}{\mathop}{egreekx}{"51}
\DeclareMathSymbol{\coprodop}{\mathop}{egreekx}{"60}
\makeatletter
\def\sum{\DOTSI\sumop\slimits@}
\def\prod{\DOTSI\prodop\slimits@}
\def\coprod{\DOTSI\coprodop\slimits@}
\makeatother
 \DeclareMathSymbol{\zeta}{\mathalpha}{egreeki}{"10}
 \DeclareMathSymbol{\iota}{\mathalpha}{egreeki}{"13}
 \DeclareMathSymbol{\kappa}{\mathalpha}{egreeki}{"14}
 \DeclareMathSymbol{\lambda}{\mathalpha}{egreeki}{"15}
 \DeclareMathSymbol{\nu}{\mathalpha}{egreeki}{"17}
 \DeclareMathSymbol{\xi}{\mathalpha}{egreeki}{"18}
 \DeclareMathSymbol{\sigma}{\mathalpha}{egreeki}{"1B}
 \DeclareMathSymbol{\varZeta}{\mathalpha}{egreeki}{"5A}
 \DeclareMathSymbol{\varIota}{\mathalpha}{egreeki}{"49}
 \DeclareMathSymbol{\varKappa}{\mathalpha}{egreeki}{"4B}
 \DeclareMathSymbol{\varLambda}{\mathalpha}{egreeki}{"03}
 \DeclareMathSymbol{\varNu}{\mathalpha}{egreeki}{"4E}
 \DeclareMathSymbol{\varXi}{\mathalpha}{egreeki}{"04}
 \DeclareMathSymbol{\varSigma}{\mathalpha}{egreeki}{"06}
 \DeclareMathSymbol{\varUpsilon}{\mathalpha}{egreeki}{"07}

\usepackage{mathdots}

\usepackage[usenames]{xcolor}
\definecolor{mybluishpurple}{RGB}{51,34,136}
\definecolor{myblue}{RGB}{136,204,238}
\definecolor{mybluishgreen}{RGB}{68,170,153}
\definecolor{mygreen}{RGB}{17,119,51}
\definecolor{mygreenishyellow}{RGB}{153,153,51}
\definecolor{myyellow}{RGB}{221,204,119}
\definecolor{myred}{RGB}{204,102,119}
\definecolor{mypurplishred}{RGB}{136,34,85}
\definecolor{myreddishpurple}{RGB}{170,68,153}
\definecolor{mygrey}{RGB}{221,221,221}

\usepackage{microtype}

\usepackage[backend=biber,mcite,subentry,citestyle=numeric-comp,bibstyle=numericbringhurst,autopunct=false,sorting=none,sortcites=false,natbib=false,maxnames=8,minnames=8,giveninits=true,block=space,hyperref=true,defernumbers=false,useprefix=true,language=british]{biblatex}

\DeclareDelimFormat{multicitedelim}{\addsemicolon\space}
\DeclareDelimFormat{postnotedelim}{\space}
\renewcommand{\bibfont}{\footnotesize}
\newcommand*{\citep}{\parencites}
\newcommand*{\citey}{\parencites*}
\renewcommand*{\cite}{\citep}
\providecommand{\href}[2]{#2}

\newcommand*{\amp}{\&}

\usepackage{graphicx}

\PassOptionsToPackage{hyphens}{url}\usepackage[hypertexnames=false]{hyperref}
\usepackage[depth=4]{bookmark}
\hypersetup{colorlinks=true,bookmarksnumbered,pdfborder={0 0 0.25},citebordercolor={0.2 0.1333 0.5333},
citecolor=mybluishpurple,linkbordercolor={0.0667 0.4667 0.2},
linkcolor=mypurplishred,urlbordercolor={0.5333 0.1333 0.3333},
urlcolor=mygreen,breaklinks=true,pdftitle={\pdftitle},pdfauthor={\pdfauthor}}
\usepackage{datetime2}
\DTMnewdatestyle{mydate}%
{
}
\DTMsetdatestyle{mydate}

\usepackage{booktabs}       

\newcommand*{\defd}{\coloneqq}
\newcommand*{\defs}{\eqqcolon}
\renewcommand*{\|}{\mathpunct{|}}
\renewcommand{\ge}{\geqslant}
\DeclarePairedDelimiter\clcl{[}{]}
\DeclarePairedDelimiter\set{\{}{\}}
\newcommand*{\p}{\mathrm{P}}
\newcommand*{\sect}{\S}
\newcommand*{\chap}{ch.}%
\newcommand*{\eqn}{eq.}%
\newcommand*{\eqns}{eqs}%
\newcommand*{\fig}{fig.}%
\newcommand*{\vs}{{vs}}
\newcommand*{\cf}{{cf.}}
\newcommand*{\etal}{{et al.}}
\newcommand*{\T}{^\intercal}
\newcommand*{\E}{\mathrm{E}}
\DeclarePairedDelimiter\expp{(}{)}
\newcommand*{\expe}{\E\expp}
\newcommand*{\expeb}{\E\clcl}

\newtheoremstyle{simple}%
{}
{}%
{\footnotesize}%
{}%
{}%
{}%
{0pt}%
{}%
\theoremstyle{simple}

\definecolor{notecolour}{RGB}{68,170,153}

\newcommand*{\widebar}[1]{{\mkern1.5mu\skew{2}\overline{\mkern-1.5mu#1\mkern-1.5mu}\mkern 1.5mu}}

\newcommand*{\tav}{\widehat} 
\newcommand*{\av}{\widebar} 
\newcommand*{\sav}{\widebar} 

\newcommand*{\ypp}{\varPi}

\newcommand*{\yXv}{\varSigma}
\newcommand*{\yRv}{\sigma}
\newcommand*{\yxv}{S}
\newcommand*{\yrv}{s}
\newcommand*{\yNv}{\nu}
\newcommand*{\yNN}{\varNu}

\newcommand*{\yx}{\bm{\yxv}}
\newcommand*{\yxs}{\sav{\yx}}
\newcommand*{\yX}{\bm{\yXv}}
\newcommand*{\yXf}{\av{\yX}}
\newcommand*{\yxxs}{\sav{\yr\yr}}
\newcommand*{\yr}{\bm{\yrv}}
\newcommand*{\yrs}{\sav{\yr}}
\newcommand*{\yR}{\bm{\yRv}}
\newcommand*{\yRf}{\av{\yR}}
\newcommand*{\yH}{\varIota}

\newcommand*{\yHa}{\varIota_\textrm{p}}
\newcommand*{\yHb}{\varIota_\textrm{s}}
\newcommand*{\yHc}{\varKappa}
\newcommand*{\yHd}{\varIota_\textrm{u}}

\newcommand*{\yf}{\bm{g}}
\newcommand*{\yc}{\bm{c}}
\newcommand*{\yL}{\bm{\lambda}}
\newcommand*{\yl}{\bm{l}}
\newcommand*{\yk}{z}
\newcommand*{\yK}{\zeta}

\newcommand*{\me}{maximum-entropy}

\title{\pdftitle}

\author{
  P.G.L. Porta Mana\\\texttt{pgl@portamana.org}
  \And
  V. Rostami\\\texttt{vrostami@uni-koeln.de}
  \AND
  E. Torre\\\texttt{emiliano.torre@yahoo.com}
    \And
Y. Roudi\\\texttt{yasser.roudi@ntnu.no}
}

\begin{document}

\maketitle

\vspace{-2em}
\begin{center}
  \footnotesize\published
\end{center}
\medskip

\begin{abstract}
  The present work shows that the \me\ method can be applied to a sample of
  neuronal recordings along two different routes: (1) apply to the sample;
  or (2) apply to a larger, unsampled neuronal population from which the
  sample is drawn, and then marginalize to the sample. These two routes
  give inequivalent results. The second route can be further generalized to
  the case where the size of the larger population is unknown. Which route
  should be chosen? Some arguments are presented in favour of the second.
  This work also presents and discusses probability formulae that relate
  states of knowledge about a population and its samples, and that may be
  useful for sampling problems in neuroscience.
\end{abstract}

\section{Introduction: \me\ and recordings of neuronal activity}
\label{sec:intro}

Suppose that we have recorded the firing activity of a hundred neurons,
sampled from a particular brain area. What are we to do with such data?
Gerstein, Perkel, Dayhoff \cite{gersteinetal1985} posed this question very
tersely (our emphasis):
\begin{quote}
  The principal conceptual problems are (1) \emph{defining cooperativity or
    functional grouping} among neurons and (2) \emph{formulating
    quantitative criteria} for recognizing and characterizing such
  cooperativity.
\end{quote}
These questions have a long history, of course; see for instance the 1966
review by Moore \etal\
\cite{mooreetal1966}. 
The neuroscientific literature has offered several mathematical definitions
of \enquote{cooperativity} or \enquote{functional grouping} and criteria to
quantify it.

One such quantitative criterion relies on the \me\ or relative-\me\ method
\cite{jaynes1957,jaynes1963,hobsonetal1973,sivia1996_r2006,meadetal1984}.
This criterion has been used in neuroscience at least since the 1990s,
applied to data recorded from brain areas as diverse as retina and motor
cortex
\cite{mackay1991,martignonetal1995,bohteetal2000,amarietal2003,schneidmanetal2006,shlensetal2006,mackeetal2009b,roudietal2009c,tkaciketal2009,gerwinnetal2009,mackeetal2011,mackeetal2011b,ganmoretal2011,granotatedgietal2013,tkaciketal2014b,moraetal2015,shimazakietal2015},
and it has been subjected to mathematical and conceptual scrutiny
\cite{tkaciketal2006,roudietal2009,roudietal2009b,barreiroetal2010,mackeetal2013,rostamietal2016_r2017}.

\enquote{Cooperativity} can be quantified and characterized with \me\
methods in several ways. The simplest way roughly proceeds along the
following steps. Consider the recorded activity of a sample of $n$ neurons.
\begin{enumerate}
\item The activity of each neuron, a continuous signal, is divided into $T$
  time bins and binarized in intensity, and thus transformed into a
  sequence of digits \enquote{$0$}s (inactive) and \enquote{$1$}s
  \cite[\cf][]{caianiello1961,caianiello1986}.

  Let the variable $\yrv_i(t)\in\set{0,1}$ denote the activity of the $i$th
  sampled neuron at time bin $t$. Collectively denote the $n$ activities
  with $\yr(t) \defd \bigl(\yrv_1(t),\dotsc,\yrv_n(t)\bigr)$. The
  population-averaged activity at that bin is
  $\yrs(t) \defd \sum_i \yrv_i(t)/n$. If we count the number of distinct
  pairs of active neurons at that bin we combinatorially find
  $\tbinom{n\yrs(t)}{2}\equiv n\yrs(t)\,[n\yrs(t)-1]/2$. There can be at
  most $\tbinom{n}{2}$ simultaneously active pairs, so the
  population-averaged pair activity is
  $\av{\yr \yr}(t) \defd \tbinom{n}{2}^{-1}\tbinom{n\yrs(t)}{2}$. With some
  combinatorics we see that the population-averaged activity of $m$-tuples
  of neurons is
  \begin{equation}
    \label{eq:products_intermsof_average}
    \av{\underbrace{\yr\dotsm\yr}_{\text{$m$ terms}}}(t)
    = \binom{n}{m}^{-1}\binom{n\yrs(t)}{m}.
  \end{equation}
  
  For brevity let us agree to simply call \enquote{activity} the average
  $\yrs$, \enquote{pair-activity} the average $\av{\yr \yr}$, and so on.

\item Construct a sequence of relative-\me\ distributions for the activity
  $\yrs$, using this sequence of constraints:
  \begin{itemize}
  \item the time average of the activity: $\tav{\yrs} \defd \sum_t\yrs(t)/T$;
  \item the time averages of the activity and of the pair-activity
    $\tav{\av{\yr \yr}} \defd \sum_t\av{\yr \yr}(t)/T$;
  \item \ldots
  \item the time averages of the activity, of the pair-activity, and so on, up
    to the $k$-activity.
  \end{itemize}
  Call the resulting distributions $p_1(\yrs), p_2(\yrs),\dotsc,p_k(\yrs)$.
  The time-bin dependence is now absent because these distributions can be
  interpreted as referring to any one of the time bins $t$, or to a new
  time bin (in the future or in the past) containing new data.

  We also have the empirical frequency distribution of the total activity,
  $f(\yrs)$, counted from the time bins.

\item Now compare the distributions above with one another and with the
  frequency distribution, using some probability-space distance like the
  relative entropy or discrimination information
  \citep{kullback1987,jaynes1963,hobson1969,hobsonetal1973}. If we find,
  say, that such distance is very high between $p_1$ and $f$, very low
  between $p_2$ and $f$, and is more or less the same between all $p_m$ and
  $f$ for $m \ge 2$, then we can say that there is a \enquote{pairwise
    cooperativity}, and that any higher-order cooperativity is just a
  reflection or consequence of the pairwise one. The reason is that the
  information from higher-order simultaneous activities did not lead to
  appreciable changes in the distribution obtained from pair activities.
\end{enumerate}
The protocol above needs to be made precise by specifying various
parameters, such as the width of the time bins or the probability
distance used.

We hurry to say that the description just given is just \emph{one} way to
quantify and characterize cooperativity and functional grouping, not
\emph{the only} way. It can surely be criticized from many points of view.
Yet, it is quantitative and bears a more precise meaning than an undefined,
vague notion of \enquote{cooperativity}. Two persons who apply this
procedure to the same data will obtain the same numbers. Different
protocols can be based on the \me\ method, for instance protocols that take
into account the activities or pair activities of specific neurons rather
than population averages, or even protocols that take into account time
dependence.

The purpose of the present work, based on the discussion and results of
\cite{portamanaetal2015}, is not to assess the merits of \me\ methods with
respect to other methods. Its main purpose is to show that there is a
problem in the particular application, sketched above, of the \me\ method
to the activity of the recorded neurons. We believe that this problem and
possible misapplication is at the root of some observations made in the
literature \cite{roudietal2009b}. This problem also extends to more
complex applications of the method, possibly excepting versions that use
\enquote{hidden} neurons
\cite{smolensky1986,kulkarnietal2007,huang2015,dunnetal2017}.

The problem is that the recorded neurons are a \emph{sample} from a larger,
unrecorded population, but the \me\ method as applied above is treating
them as isolated from the rest of the brain. Hence the results it provides
cannot rightfully be extrapolated. We will give a mathematical proof of
this. Let us first analyse this issue in more detail.

Suppose that the neurons were recorded with electrodes covering an area of
some square millimetres \cite[\cf][]{berenyietal2014}. This recording is a
sample of the activity of the neuronal population under the recording
device, a population that can amount to tens of thousands of neurons
\cite{abeles1991}. We could even consider the recorded neurons as a sample
of a brain area more extended than the recording device.

The characterization of the cooperativity of the recorded sample would have
little meaning if we did not expect its results to generalize to a larger,
unrecorded population -- at the very least the one under the recording
device. In other words, we expect that the conclusions drawn with the \me\
methods about the sampled neurons should somehow extrapolate to unrecorded
neurons in some larger area, from which the recorded neurons were sampled.
In statistical terms we are assuming that the recorded neurons are a
\emph{representative sample}\footnote{Note that the \textsc{iso} standard
  \cite[\sect~3.1.14]{iso2014} states \enquote{The notion of representative
    sample is fraught with controversy, with some survey practitioners
    rejecting the term altogether}. Here we intend this notion in the
  Meaning~2 of Kruskal \amp\ Mosteller \cite{kruskaletal1979b} or
  Meanings~8 and~9 of \cite{kruskaletal1979c}.} of some larger neuronal
population. Probability theory tells us how to make inferences from a
sample to the larger population from which it is sampled \parentext{see
  references below}.

We can apply the \me\ method to the sample, as described in the above
protocol, to generate probability distributions for the activity of the
sample. But, given that our sample is representative of a larger
population, we can also apply the \me\ method to the larger (unrecorded)
population. The constraints are the same: namely the time averages of the
sampled data -- in fact they constitute representative data about the
larger population as well. The method thus yields a probability
distribution for the larger population, and the distribution for the sample
is then obtained by marginalization from that. The problem is that
\emph{the distributions obtained from these two applications differ}. Which
choice is most meaningful?

In this work we develop the second way \cite[sketched
in][\sect~3]{portamanaetal2015} of applying the \me\ method, at the level
of the larger population, and show that its results differ from the
application at the sample level. We also consider the case where the size
of the larger population is unknown.

To apply the \me\ method to the larger, unsampled population, it is
necessary to use probability relations relevant to sampling
\cites{ghoshetal1997}[parts~I,
VI]{freedmanetal1978_r2007}[\chap~3]{jaynes1994_r2003}{portamanaetal2015}.
The relations we present are well-known in survey sampling and in the
pedagogic problem of drawing from an urn without replacement, yet they are
somewhat hard to find explicitly written in the neuroscientific literature.
We present and discuss them in the next section. A minor purpose of this
paper is to make these relations more widely known, because they can be
useful independently of \me\ methods.

The notation and terminology in the present work follow \textsc{iso} and
\textsc{ansi} standards \citep{iso1993,ieee1993,nist1995,iso2006,iso2006b}
but for the use of the comma \enquote{,} to denote logical conjunction.
Probability notation follows Jaynes \citep{jaynes1994_r2003}. By
\enquote{probability} we mean a degree of belief which \enquote{would be
  agreed by all rational men if there were any rational men}
\cite{good1966}.

\section{Probability relations between population and  sample}
\label{sec:prob_samples}

We have already introduced the notation for the sample neurons. We
introduce an analogous notation for the $\yNv$ neurons constituting the
larger population, but using the corresponding Greek letters:
$\yRv_{\iota}(t)$ is the activity of the $\iota$th neuron at time bin $t$,
$\yRf(t) \defd \sum_{\iota} \yRv_{\iota}(t)/\yNv$ is the activity at that
bin averaged over the larger population, and so on.

The probability relations between sample and larger population are valid at
every time bin. As we mentioned above, the \me\ distribution refers to any
time bin or to a new bin. For these reasons we will now omit the time-bin
argument \enquote{(t)} from our expressions. 

Probabilities refer to statements about the quantities we observe. We use
the standard notation:
\begin{equation}
  \label{eq:notation_statements}
  \begin{aligned}
    &\text{\enquote{$\yXv_{\iota} = \yRv_{\iota}$} 
      stands for \enquote{the activity of the $\iota$th neuron is $\yRv_{\iota}$}},
    \\
    &\text{\enquote{$\yXf = \yRf$} 
      stands for \enquote{the (population-averaged) activity of the neurons is $\yRf$}},
    \\
    &\text{\enquote{$\yxv_i = \yrv_i$} 
      stands for \enquote{the activity of the $i$th sample neuron is $\yrv_i$}},
  \end{aligned}
\end{equation}
and similarly for other quantities.

If $\yHc$ denotes our state of knowledge -- the evidence and assumptions
backing our probability assignments -- our uncertainty about the full
activity of the larger population is expressed by the joint probability
distribution
\begin{equation}
  \label{eq:joint_plaus}
  \p(\yXv_1=\yRv_1, \yXv_2=\yRv_2, \dotsc, \yXv_\yNv=\yRv_\yNv \| \yHc)
  \quad\text{or}\quad
\p(\yX =\yR \| \yHc), \quad \yR \in \set{0,1}^\yNv.
\end{equation}
Our uncertainty about the state of the sample is likewise expressed by
\begin{equation}
  \label{eq:sample_plaus}
  \p(\yxv_1=\yrv_1, \yxv_2=\yrv_2, \dotsc, \yxv_n=\yrv_n \| \yHc) \quad\text{or}\quad
\p(\yx =\yr \| \yHc), \quad \yr \in \set{0,1}^n.
\end{equation}

\bigskip

The theory of statistical sampling is covered in many excellent texts, for
example Ghosh \amp\ Meeden \citey{ghoshetal1997} or Freedman, Pisani, \amp\
Purves \citey[parts~I, VI]{freedmanetal1978_r2007}; a summary can be found
in Jaynes
\citey[\chap~3]{jaynes1994_r2003}.

We need to make an initial probability assignment for the state of the full
population before any experimental observations are made. This initial
assignment will be modified by our experimental observations, and these can
involve just a sample of the population. Our state of knowledge and initial
probability assignment should reflect that samples are somehow
representative of the whole population.

In this state of knowledge, denoted $\yH$, we know that the neurons in the
population are biologically or functionally similar, for example in
morphology or the kind of input or output they receive or give. But we are
completely ignorant about the physical details of the individual neurons.
Our ignorance is therefore symmetric under permutations of neuron
identities. This ignorance is represented by a probability distribution
that is symmetric under permutations of neuron identities; such a
distribution is usually called \emph{finitely exchangeable}
\cites{ericson1969}[\chap~1]{ghoshetal1997}. We stress that this
probability assignment is just an expression of the symmetry of our
\emph{ignorance} about the state of the population, not an expression of
some biologic or physical symmetry or identity of the neurons.

The \emph{representation theorem for finite exchangeability} states that,
in the state of knowledge $\yH$, the symmetric distribution for the full
activity is completely determined by the distribution for its
population-average:
\begin{equation}
  \label{eq:joint_plaus_N_homog}
  \p(\yX = \yR \|  \yH) \equiv
  \sum_{\yRf}\p(\yX = \yR \| \yXf = \yRf, \yH)\,
  \p(\yXf=\yRf \| \yH) =
  \binom{\yNv}{\yNv\yRf}^{-1} \p(\yXf=\yRf \| \yH).
\end{equation}
The equivalence on the left is just an application of the law of total
probability; the equality on the right is the statement of the theorem.
This result is intuitive: owing to symmetry, we must assign equal
probabilities to all $\tbinom{\yNv}{\yNv\yRf}$ activity vectors with
$\yNv\yRf$ active neurons; the probability of each activity vector is
therefore given by that of the average activity divided by the number of
possible vector values. Proof of this theorem and generalizations to
non-binary and continuum cases are given by de~Finetti
\cite{definetti1959b}, Kendall \cite{kendall1967}, Ericson
\cite{ericson1976}, Diaconis \amp\ Freedman
\cite{diaconis1977,diaconisetal1980}, Heath \amp\ Sudderth
\cite{heathetal1976}.

Our uncertainties about the full population and the sample are connected
via the conditional probability
\begin{equation}
  \label{eq:conditional_hypergeometric}
  \p(\yxs = \yrs \|\yXf=\yRf, \yH)=
  \binom{n}{n\yrs}\binom{\yNv-n}{\yNv \yRf-n\yrs}\binom{\yNv}{\yNv \yRf}^{-1}
  \defs \ypp(\yrs\|\yRf),
\end{equation}
which is a hypergeometric distribution, typical of \enquote{drawing without
  replacement} problems. The combinatorial proof of this expression is in
fact the same as for this class of problems
\cites[\chap~3]{jaynes1994_r2003}[\sect~4.8.3]{ross1976_r2010}[\sect~II.6]{feller1950_r1968}.

Using the conditional probability above we obtain the probability for the
activity of the sample:
\begin{gather}
  \label{eq:subpop_average}
  \p(\yxs=\yrs \| \yH) = \sum_{\yRf}
  \p(\yxs = \yrs \|\yXf=\yRf, \yH)\,
  \p(\yXf=\yRf \| \yH)
  = 
  \sum_{\yRf}
  \ypp(\yrs\|\yRf)\,
  \p(\yXf=\yRf \| \yH).
\end{gather}
It should be proved that the probability distribution for the full activity
of the sample is also symmetric and completely determined by the
distribution of its population-averaged activity:
\begin{equation}
  \label{eq:marginal}
  \p(\yx = \yr \| \yH) = \binom{n}{n\yrs}^{-1} \p(\yxs=\yrs \| \yH).
\end{equation}
This is intuitively clear: our initial symmetric ignorance should also
apply to the sample. The distribution for the
sample~\eqref{eq:subpop_average} indeed satisfies the same representation
theorem~\eqref{eq:joint_plaus_N_homog} as the distribution for the full
population.

The conditional probability
$\p(\yxs = \yrs \|\yXf=\yRf, \yH) \equiv \ypp(\yrs \|\yRf)$, besides
relating the distributions for the population and sample activities via
marginalization, also allows us to express the expectation value of any
function of the sample activity, $g(\yrs)$, in terms of the distribution
for the full population, as follows:
\begin{multline}
  \label{eq:pullback_P}
  \expe{g\|I}
  \equiv
  \sum_{\yrs} g(\yrs)\,\p(\yxs=\yrs \| \yH)
  =
  \sum_{\yrs} g(\yrs) \sum_{\yRf} \ypp(\yrs\|\yRf)\,\p(\yXf=\yRf \| \yH)
  ={}\\
  \sum_{\yRf} \biggl[ \sum_{\yrs} g(\yrs)  \ypp(\yrs\|\yRf) \biggr]\,
  \p(\yXf=\yRf \| \yH),
\end{multline}
where the second step uses \eqn~\eqref{eq:subpop_average}. The last
expression shows that the expectation of the function $g(\yrs)$ is equal to
the expectation of the function
$g^*(\yRf) \defd \sum_{\yrs} g(\yrs)\,\ypp(\yrs\|\yRf)$.

\bigskip

The final expression in \eqn~\eqref{eq:pullback_P} is important for our
\me\ application: the requirement that the function $g$, defined for the
sample, have a  value $c$ obtained from observed data,
\emph{translates into a linear constraint for the distribution of the full
  population}:
\begin{equation}
  \label{eq:constraint_extended}
  c = \expe{g \| \yH} \equiv \sum_{\yRf} \biggl[ \sum_{\yrs} g(\yrs)  \ypp(\yrs\|\yRf) \biggr]\,
  \p(\yXf=\yRf \| \yH).
\end{equation}

In particular, when the function $g$ is the $m$-activity of the sample,
$g(\yrs) = \sav{\yr\dotso\yr} \equiv \binom{n\yrs}{m}/\binom{n}{m}$, we
find
\begin{equation}
  \label{eq:expe_products}
  \expe{\sav{\underbrace{\yr \dotsm \yr}_{\text{$m$ factors}}} \| \yH}
\equiv
    \sum_{\yrs} 
    \binom{n}{m}^{-1}
    \binom{n \yrs}{m}\, \p(\yxs=\yrs \| \yH)
    = \binom{\yNv}{m}^{-1}
    \sum_{\yRf} 
    \binom{\yNv \yRf}{m}\, \p(\yXf=\yRf \| \yH)
\equiv    \expe{\av{\underbrace{\yR \dotsm \yR}_{\text{$m$ factors}}} \| \yH},
\end{equation}
that is, \emph{the expected values of the $m$-activities of the sample and
  of the full population are equal}. The proof of the middle equality uses
the expression for the $m$th factorial moment of the hypergeometric
distribution and can be found in \textcite{potts1953}. Similar relations
can be found for the raw moments $\expe{\yrs^m}$ and $\expe{\yRf^m}$, which
can be written in terms of the product expectations using
\eqn~\eqref{eq:products_intermsof_average}.

Thus, in a \me\ application, when we require the expectation of the
$m$-activity of a sample to have a particular value, we are also
requiring  the expectation of the $m$-activity of the full population to
have the same value.

\bigskip

\begin{figure}[!b]
\centering
\includegraphics[width=0.9\linewidth]{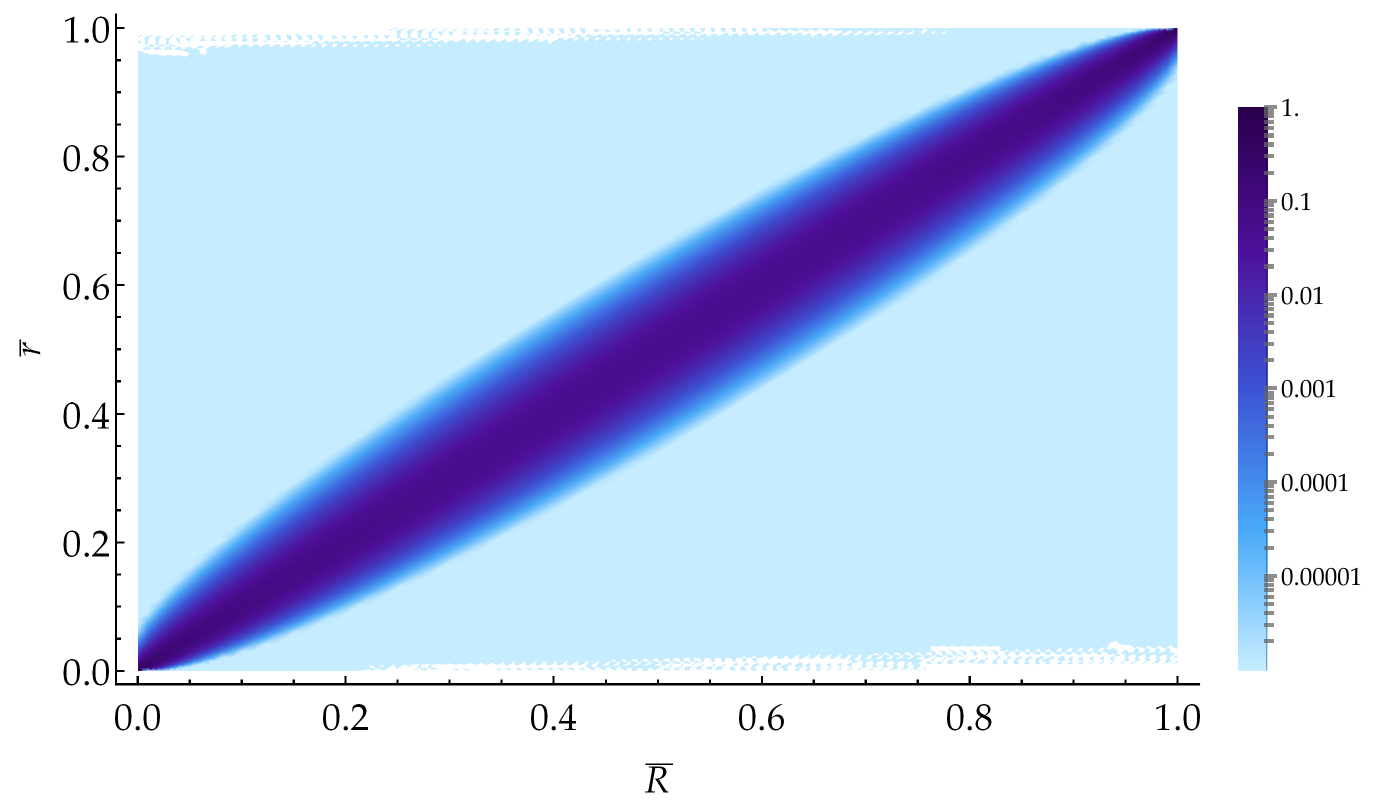}%
\caption{Log-density plot of the hypergeometric distribution
  $
\ypp(\yrs\|\yRf) \defd  \raisebox{0pt}[0pt][1ex]{}\binom{n}{n\yrs}\binom{\yNv-n}{\yNv \yRf-n\yrs}\binom{\yNv}{\yNv \yRf}^{-1}$ for $\yNv=5000$, $n=200$. (Band artifacts may appear in the
  colourbar depending on your \textsc{pdf} viewer.)}
\label{fig:hypergeom_proj}
\end{figure}
These expectation equalities between sample and full population should not
be surprising: we intuitively \emph{expect} that the proportion of coloured
balls sampled from an urn should be roughly equal to the proportion of
coloured ball contained in the urn. The formulae in the present section
formalize and mathematically express our intuition. The hypergeometric
distribution $\ypp(\yrs\|\yRf)$ plays an important role in this
formalization. A look at its plot, \fig~\ref{fig:hypergeom_proj}, reveals
that it is a sort of \enquote{fuzzy identity transformation}, or fuzzy
Kronecker delta, between the $\yRf$-space $\set{0,\dotsc,\yNv}$ and
$\yrs$-space $\set{0,\dotsc,n}$. From \eqn~\eqref{eq:marginal} we thus have
that
\begin{equation}
  \label{eq:roughly_equal_nN}
  \p(\yxs=a \|\yH) \approx \p(\yXf=a \|\yH),\qquad
\expeb{g(\yrs) \|\yH} \approx \expeb{g(\yRf) \|\yH},
\end{equation}
where $g$ is any smooth function defined on $\clcl{0,1}$. These approximate
equalities express the intuitive fact that \emph{our uncertainty about the
  sample is representative of our uncertainty about the population and
  about other samples}, and vice versa. When $n=\yNv$, $\ypp(\yrs\|\yRf)$
becomes the identity matrix and the approximate equalities above become
exact -- of course, since we have sampled the full population.

But the approximate equalities above may miss important features of the two
probability distributions. In the next section we will in fact emphasize
their differences. If the distribution for the population average $\yRf$ is
bimodal, for example, the bimodality can be lost in the distribution for
the sample average $\yrs$, owing to the coarsening effect of
$\ypp(\yrs\|\yRf)$.


\section{Maximum-entropy: sample level \vs\ full-population level}
\label{sec:specific_initial_probability}

In the previous section we have seen that observations about a sample can
be used as constraints on the distribution for the activity of the full
population. Let us use such constraints with the \me\ method. Suppose that
we want to constrain $m$ functions of the sample activity, vectorially
written $\yf \defd (g_1,\dotsc,g_m)$, to $m$ values
$\yc \defd (c_1,\dotsc,c_m)$. These functions are typically $k$-activities
$\sav{\yr\dotso \yr}$, and the values are typically the time averages of
the observed sample, as discussed in \sect~\ref{sec:intro}:
$\yc = \sum_t \yf[\yrs(t)]/T$.

Let us apply the relative-\me\ method \cite{sivia1996_r2006,meadetal1984}
directly to sampled neurons; denote this approach by $\yHb$. Then we apply
the method to the full population of neurons, most of which are unsampled;
denote this approach by $\yHa$.

Applied directly to the sampled neurons, the method yields the distribution
\begin{equation}
  \label{eq:app_maxent_sample}
  \p(\yxs=\yrs \|\yHb)
  =\frac{1}{\yk(\yl)}\,
  \binom{n}{n\yrs}
  \exp[\yl\T \yf(\yrs)]
\end{equation}
where $\yk(\yl)$ is a normalization constant. The binomial in front of the
exponential appears because we must account for the multiplicity by which
the population-average activity $\yrs$ can be realized: $\yrs=0$ can be
realized in only one way (all neurons inactive), $\yrs=1/n$ can be realized
in $n$ ways (one active neuron out of $n$), and so on. This term is analogous to
the \enquote{density of states} in front of the Boltzmann factor in
statistical mechanics \cite[\chap~16]{callen1960_r1985}. The $m$ Lagrange
multipliers $\yl\defd (l_1,\dotsc,l_m)$ must satisfy the $m$ constraint
equations
\begin{equation}
  \label{eq:app_maxent_sample_constraints}
  c_k = \expe{g_k \|\yHb} \equiv
  \frac{1}{\yk(\yl)}\sum_{\yrs}  g_k(\yrs) \binom{n}{n\yrs}
  \exp[\yl\T \yf(\yrs)], \qquad k=1,\dotsc,m.
\end{equation}

Applied to the full population, using the constraint
expression~\eqref{eq:constraint_extended} derived in the previous section,
the method yields the distribution for the full-population activity
\begin{equation}
  \label{eq:app_maxent_pop}
  \p(\yXf = \yRf \| \yHa)  = \frac{1}{\yK(\yL)}\,
  \binom{\yNv}{\yNv \yRf}\,\exp\Bigl[\yL\T
  \sum_{\yrs} \yf(\yrs)\ypp(\yrs\|\yRf)\Bigr].
\end{equation}
The $m$ Lagrange multipliers $\yL\defd (\lambda_1,\dotsc,\lambda_m)$ must
satisfy the $m$ constraint equations
\begin{equation}
  \label{eq:app_maxent_pop_constraints}
  c_k = \expe{g_k \|\yHa}\equiv
  \frac{1}{\yK(\yL)}\sum_{\yrs} \sum_{\yRf} g_k(\yrs) \ypp(\yrs \|\yRf)
\,
  \binom{\yNv}{\yNv \yRf}\,\exp\Bigl[\yL\T
  \sum_{\yrs} \yf(\yrs)\ypp(\yrs\|\yRf)\Bigr],
  \qquad k=1,\dotsc,m.
\end{equation}

We obtain the distribution for the sample activity by marginalization, using
\eqn~\eqref{eq:marginal}:
\begin{equation}
  \label{eq:app_maxent_pop_marg}
  \p(\yxs = \yrs \| \yHa)  = \frac{1}{\yK(\yL)}\, 
  \sum_{\yRf} \ypp(\yrs \|\yRf)
\,
  \binom{\yNv}{\yNv \yRf}\,\exp\Bigl[\yL\T
  \sum_{\yrs} \yf(\yrs)\ypp(\yrs\|\yRf)\Bigr].
\end{equation}

The distributions for the sample activity,
\eqns~\eqref{eq:app_maxent_pop_marg} and \eqref{eq:app_maxent_sample},
obtained with the two approaches $\yHb$ and $\yHa$, are different. From the
discussion in the previous section we expect them to be vaguely similar;
yet they cannot be exactly equal, because their equality would require the
$2m$ quantities $\yL$ and $\yl$ to satisfy the constraint equations
\eqref{eq:app_maxent_pop_constraints} and
\eqref{eq:app_maxent_sample_constraints}, and in addition also the $n$
equations $\p(\yxs = \yrs \| \yHa) = \p(\yxs = \yrs \| \yHb)$,
$\yrs=1/n,\dotsc,1$ (one equation is taken care of by the normalization of
the distributions). We would have a set of $2m+n$ equations in $2m$
unknowns.

Hence, \emph{the applications of \me\ at the sample level and at the
  full-population level are inequivalent}. They lead to numerically
different distributions for the sample activity $\yr$.

%
\begin{figure}[!t]
\centering
\includegraphics[width=0.9\linewidth]{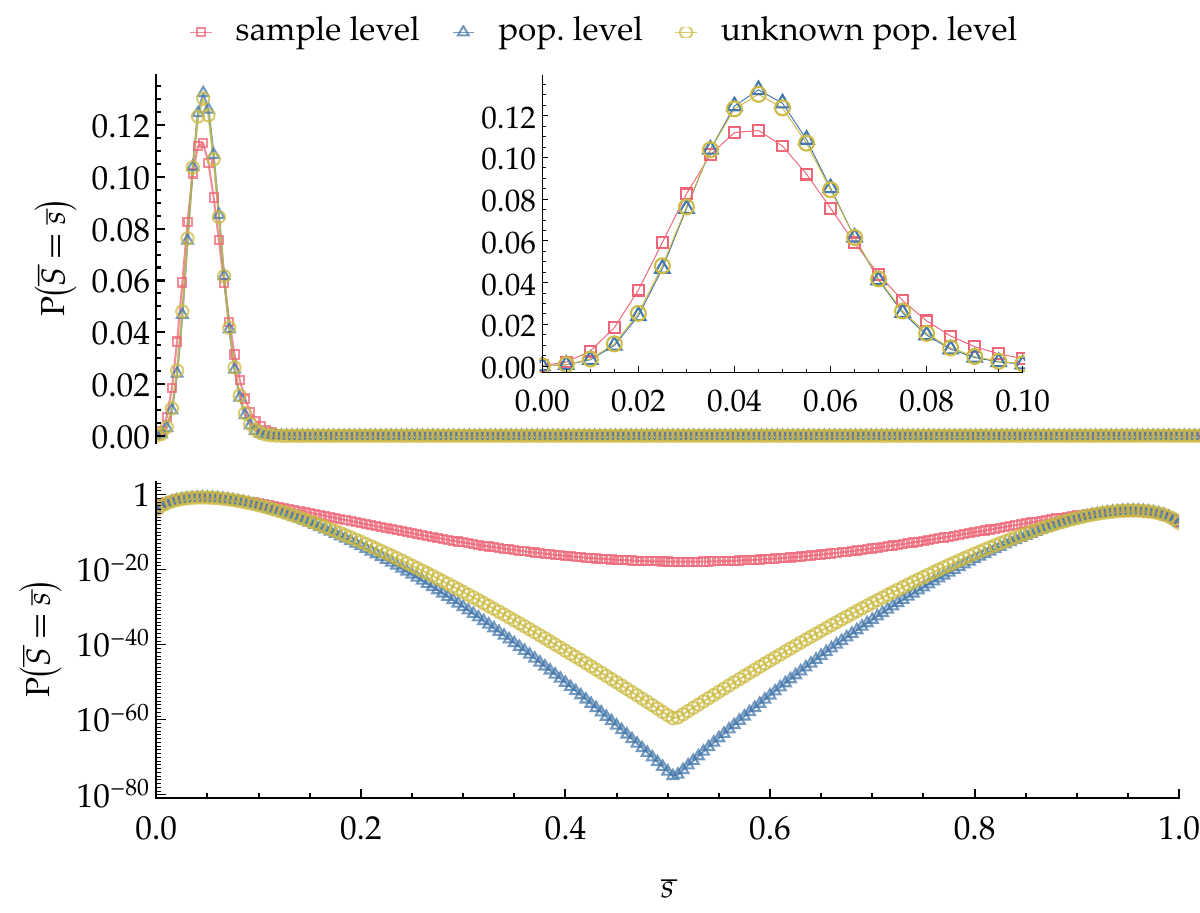}%
\caption{Linear and log-plots of $\p(\yxs = \yrs)$ for a sample of $n=200$
  and constraints as in \eqn~\eqref{eq:constraints}, constructed by:
  \textcolor{myred}{\textbf{red squares:}} \me\ at the sample level,
  \eqn~\eqref{eq:app_maxent_sample};
  \textcolor{mybluishpurple}{\textbf{blue triangles:}} \me\ at the
  population level, \eqn~\eqref{eq:app_maxent_pop_marg} with
  $\yNv=10\,000$, followed by sample marginalization;
  \textcolor{myyellow}{\textbf{yellow circles:}} \me\ at the population
  level with unknown population size,
  \eqn~\eqref{eq:app_maxent_pop_marg_unknown_N}, according to the
  distribution~\eqref{eq:pdf_popsize} for the population.}
\label{fig:all_three}
\end{figure}
The distribution obtained at the sample level will show different
features from the one obtained at the population level, like displaced or
additional modes or particular tail behaviour. We show an example of this
discrepancy in \fig~\ref{fig:all_three}, for
$\yNv=10\,000$, $n=200$, and the two constraints
\begin{equation}
  \label{eq:constraints}
  \expe{\yrs} = 0.0478,\qquad
  \expe{\yxxs} = 0.00257,
\end{equation}
which come from the actual recording of circa 200 neurons from macaque
motor cortex \cite{rostamietal2016_r2017}. The distribution obtained at the
population level (blue triangles) has a higher and displaced mode and a quite
different behaviour for activities around $0.5$ than the distribution
obtained at the sample level (red squares).

\bigskip

In our discussion we have so far assumed the size $\yNv$ of the larger
population to be known. This is rarely the case, however. We usually are
uncertain about $\yNv$ and can only guess its order of magnitude. In such a
state of knowledge $\yHd$ our ignorance about the possible value of $\yNv$
is expressed by a probability distribution $\p(\yNN=\yNv \|\yHd)=h(\yNv)$,
and the marginal distribution for the sample
activity~\eqref{eq:app_maxent_pop_marg} is modified, by the law of total
probability, to
\begin{multline}
  \label{eq:app_maxent_pop_marg_unknown_N}
  \p(\yxs = \yrs \| \yHd)  =
  \sum_{\yNv} \p(\yxs = \yrs \| \yNN=\yNv,\yHd) \,
  \p(\yNN=\yNv \|\yHd)
  ={}\\[-\jot]
  \sum_\yNv \biggl\{\frac{1}{\yK(\yL_\yNv)}\, 
  \sum_{\yRf} \ypp_\yNv(\yrs \|\yRf)
\,
  \tbinom{\yNv}{\yNv \yRf}\,\exp\Bigl[{\yL_\yNv}\T
  \sum_{\yrs} \yf(\yrs)\ypp_\yNv(\yrs\|\yRf)\Bigr]\biggr\}
  \,h(\yNv),
\end{multline}
where the Lagrange multipliers $\yL_{\yNv}$ and the summation range for
$\yRf$ depend on $\yNv$.

As a proof of concept, \fig~\ref{fig:all_three} also shows such a distribution
(yellow circles) for the same constraints as above, and a probability
distribution for $\yNv$ inspired by Jeffreys
\cite[\sect~4.8]{jeffreys1939_r1983}:
\begin{equation}
  \label{eq:pdf_popsize}
  h(\yNv) \propto 1/\yNv, \qquad
  \yNv \in\set{1\,000,\; 2\,000,\; \dotsc,\;10\,000}.
\end{equation}

\section{Discussion}
\label{sec:discussion}

The purpose of the present work was to point out and show, in a simple
set-up, that the \me\ method can be applied to recorded neuronal data in a
way that accounts for the larger population from which the data are
sampled, \eqns~\eqref{eq:app_maxent_pop}--\eqref{eq:app_maxent_pop_marg}.
This application leads to results that differ from the standard application
which only considers the sample in isolation,
\eqns~\eqref{eq:app_maxent_sample}--\eqref{eq:app_maxent_sample_constraints}.
We gave a numerical example of this difference. We have also shown how to
extend the new application when the size of the larger population is
unknown, \eqn~\eqref{eq:app_maxent_pop_marg_unknown_N}.

The latter formula, in particular, shows that the standard way of applying
\me\ 
implicitly assumes that \emph{no} larger population exists beyond the
recorded sample of neurons. One could in fact object to the application at
the population level, and say that the traditional way of applying \me,
\eqn~\eqref{eq:app_maxent_sample}, yields different results because it does
not make assumptions about the size $\yNv$ of a possibly existing larger
population. Such a state of uncertainty, however, is correctly formalized
according to the laws of probability by introducing a probability
distribution for $\yNv$, and is expressed by
\eqn~\eqref{eq:app_maxent_pop_marg_unknown_N}. This expression cannot
generally be equal to~\eqref{eq:app_maxent_sample} unless the distribution
for $\yNv$ gives unit probability to $\yNv=n$; that is, unless the sample
\emph{is} the full population, and no larger population exists.

The standard \me\ approach therefore assumes that the recorded neurons
constitute a special subnetwork, isolated from the larger network of
neurons in which it is embedded, and which was also present under the
recording device. This assumption is unrealistic. The \me\ approach at the
population level does not make such assumption and is therefore preferable.
It may reveal features in a data set that were unnoticed by the standard
\me\ approach.


The difference in the resulting distributions between the applications at
the sample and at the population levels appears in the use of Boltzmann
machines with hidden units \cite{lerouxetal2008}, although by a different
conceptual route. It also appears in statistical mechanics: if a system is
statistically described by a \me\ Gibbs state, its subsystems cannot be
described by a Gibbs state \cite{maesetal1999}. A somewhat similar
situation also appears in the statistical description of the final state of
a non-equilibrium process starting and ending in two equilibrium states: we
can describe our knowledge about the final state either by (1) a Gibbs
distribution, calculated from the final equilibrium macrovariables, or (2)
by the distribution obtained from the Liouville evolution of the Gibbs
distribution assigned to the initial state. The two distributions differ
(even though the final \emph{physical} state is obviously exactly the same
\cite[\sect~4]{jaynes1985d_r1993}), and the second allows us to make
sharper predictions about the final physical state thanks to our knowledge
of its preceding dynamics. In this example, though, both distributions are
usually extremely sharp and practically lead to the same predictions. In
neuroscientific applications, the difference in predictions of the sample
\vs\ full-population applications can instead be very relevant.

The idea of the new application leads in fact to more questions. For
instance:
\begin{itemize}
\item Do the standard and new applications lead to different or contrasting
  conclusions about \enquote{cooperativity}, when applied to real data
  sets?
\item How to extend the new application to the \enquote{inhomogeneous} case
  \cite{schneidmanetal2006,shlensetal2006,roudietal2009b}, in which
  expectations for individual neurons or groups of neurons are constrained?
\item What is the mathematical relation between the new application and
  \me\ models with hidden neurons
  \cite{smolensky1986,kulkarnietal2007,huang2015,dunnetal2017}?
\end{itemize}
Owing to space limitations we must leave a thorough investigation of these
questions to future work.

Finally, we would like to point out the usefulness and importance of the
probability formulae that relate our states of knowledge about a population
and its samples, presented in \sect~\ref{sec:prob_samples}. This kind of
formulae is essential in neuroscience, where we try to understand
properties of extended brain regions from partial observations. The
formulae presented here reflect a simple, symmetric state of ignorance.
More work is needed \cite[\cf][]{levinaetal2017} to extend these formulae
to account for finer knowledge of the cerebral cortex and its network
properties.




\subsubsection*{Acknowledgements}

PGLPM thanks Mari \amp\ Miri for continuous
encouragement and affection; Buster Keaton for filling life with awe and
inspiration; the developers and maintainers of \LaTeX, Emacs, AUC\TeX, Open
Science Framework, Python, Inkscape, Sci-Hub for making a free and
unfiltered scientific exchange possible.


\renewcommand*{\bibname}{References}
\defbibheading{bibliography}[\bibname]{\section*{#1}\addcontentsline{toc}{section}{#1}
}






\newcommand*{\citein}[2][]{\textnormal{\textcite[#1]{#2}}
}
\newcommand*{\citebi}[2][]{ref.\ \citep[#1]{#2}
}
\newcommand*{\subtitleproc}[1]{}
\newcommand*{\chapb}{ch.}

\printbibliography

\end{document}